\documentstyle[aps,12pt,floats]{revtex}

\tightenlines  

\begin{document}
  \renewcommand{\floatpagefraction}{0.9}
\input{epsf}

\draft

\author{%
  Paul C. Dulany\thanks{dulany@quark.umd.edu}\ \ and S. J. Wallace
}

\address{%
  Department of Physics and Center for Theoretical Physics,\\
  University of Maryland,
  College Park, Maryland 20742-4111
}

\title{
  Relativistic three-body bound states and \\
  the reduction from four to three dimensions\thanks{From a
  dissertation to be submitted to the Graduate School, University of
  Maryland, by Paul~C.~Dulany in partial fulfillment of the
  requirements for the Ph.D.~degree in Physics.}
}

\date{
  Submitted to Phys.\ Rev.\ C February 14, 1997)\\
  (Published December 1, 1997 [Phys.\ Rev.\ C {\bf 56}, 2992]
}

\maketitle

\begin{abstract}

Beginning with an effective field theory based upon meson exchange,
the Bethe-Salpeter equation for the three-particle propagator
(six-point function) is derived.  Using the one-boson-exchange form of
the kernel, this equation is then analyzed using time-ordered
perturbation theory, and a three-dimensional equation for the
propagator is developed.  The propagator consists of a pre-factor in
which the relative energies are fixed by the initial state of the
particles, an intermediate part in which only global propagation of
the particles occurs, and a post-factor in which relative energies are
fixed by the final state of the particles.  The pre- and post-factors
are necessary in order to account for the transition from states where
particles are off their mass shell to states described by the global
propagator with all of the particle energies on shell.  The pole
structure of the intermediate part of the propagator is used to
determine the equation for the three-body bound state:  a
Schr{\"o}dinger-like relativistic equation with a single, global
Green's function.  The role of the pre- and post-factors in the
relativistic dynamics is to incorporate the poles of the breakup
channels in the initial and final states.  The derivation of this
equation by integrating over the relative times rather than via a
constraint on relative momenta allows the inclusion of retardation and
dynamical boost corrections without introducing unphysical
singularities.

\end{abstract}


\section{Introduction}

Faddeev's three-body work of 1960 \cite{faddeev:spjept-12-1014},
combined with the earlier two-body work of Bethe and Salpeter, and
Gell-Mann and Low\cite{salpeter:pr-84-1232,gell-mann:pr-84-350},
produced significant interest within the nuclear and particle physics
communities in solving the fully relativistic three-body problem 
  \cite{taylor:ncs-1-857,stojanov:pl-13-76,shelest:pl-13-253,%
  taylor:pr-150-1321,broido:rpp-32-493,freedman:nc-43-258,%
  tucciarone:nc-41-204}.  
Calculations based upon the full four-dimensional theory, however,
have proved very difficult, and have only recently been performed by
Rupp and Tjon using separable interactions \cite{rupp:prc-37-1729}.

In the meantime, three-dimensional nonrelativistic calculations based
upon the Schr{\"o}\-ding\-er equation have progressed significantly.
Using $NN$ potentials which provide a good description of two-body
data, recent calculations have been performed to a precision of 10~keV
\cite{machleidt:prc-53-R1483}.  
Given the successes of these nonrelativistic calculations in obtaining
excellent precision, discrepancies with experiment of order 10~keV
or larger are due to inaccuracies in the theoretical input rather than
uncertainties in the calculations.  The main ``missing physics'' in
these calculations are three-body forces and relativistic effects.
A recent calculation found the triton to be underbound by roughly
480--860~keV \cite{machleidt:prc-53-R1483}.  A recent assessment of
the triton potential energy suggested that a consistent relativistic
calculation could account for as much as 300 keV of repulsion
\cite[and references therein]{friar:ATPE-96}; however, recent
relativistic calculations using the Blankenbecler-Sugar formalism and
the CD-Bonn potential showed an {\em increased\/} binding of 200 keV
\cite{machleidt:prc-53-R1483}.  The size of the relativistic effects
determines the amount of three-body interaction that would therefore
be required.  A consistent relativistic three-nucleon calculation is
needed in order to permit some understanding of the respective roles
of relativistic effects and three-body forces in nuclear binding.  

In view of the successes of the nonrelativistic calculations and the
need for relativistic calculations, many three-dimensional reductions
of the relativistic four-dimensional Bethe-Salpeter equation have been
made following the two-body work of Blankenbecler and Sugar, and
Logunov and Tavkhelidze
  \cite{blankenbecler:pr-142-1051,logunov:cn-29-380,%
  alessandrini:pr-139-B167,ahmadzadeh:pr-147-1111,%
  kvinikhidze:rtp3d-71,stadler:rctbe-96}.
This approach, often called the quasipotential approach, consists of
replacing the Bethe-Salpeter equation with a set of two coupled
equations.  These equations involve two new functions, the
quasipotential $W$ and the quasipotential Green's function
$G_{\text{QP}}$.  We may choose one of these functions arbitrarily.
The requirement that these equations be equivalent to the
Bethe-Salpeter equation then fixes the other function.
Traditionally, the quasipotential Green's function $G_{\text{QP}}$ is
chosen to contain a Dirac $\delta$~function constraint which reduces
the dimensionality of one of the new equations.  Difficulties with
this procedure arise when the equation for the quasipotential $W$ is
truncated.  What form these difficulties take is dependent upon the
form of the constraint used in the quasipotential Green's function.
For example, in the Gross equation \cite{gross:prc-45-2094} this
truncation introduces into the wave function unphysical singularities
which must be removed by hand.  In the instant formalism
\cite{devine:prc-48-R973}, singularities arise when attempting the
dynamical boost of the wave function.  Although they take different
forms, the root cause of these difficulties is the $\delta$~function
constraint combined with the truncation of the quasipotential.
Given the difficulties with solving the full four-dimensional
equation, the success of the nonrelativistic three-dimensional
calculations, and the fundamental problems in the quasipotential
approach, a different technique for dimensional reduction seems
warranted.

In this paper we develop an approach to the relativistic three-body
problem, with an emphasis on three-body bound states.  We start from
the Bethe-Salpeter equation for the full three-body Green's function
in momentum space.  For definiteness, we use scalar meson exchange as
a model for the interaction; extension to other forms of
interactions is straightforward.  Negative energy states are omitted
as the dominant physics is obtained from positive energy states.

We perform a Fourier transformation of the zeroth component of all
internal momenta to relative-time variables, and carry out the
relative-time integrations, which has the effect of transforming each
Feynman graph into several time-ordered graphs.  These time-ordered
graphs are three-dimensional in nature.  Rearranging and summing
graphs produces an expression for the full, four-dimensional,
three-body Green's function in which all of the internal variables are
three dimensional.  It follows that the bound-state equation is three
dimensional as well.  We find that the propagator consists of a
pre-factor in which the relative energies are fixed by the initial
state of the particles, an intermediate part in which only global
propagation of the particles occurs, and a post-factor in which
relative energies are fixed by the final state of the particles.  The
pre- and post-factors are necessary in order to account for the
transition from states where particles are off their mass shell to
states described by the global propagator with all of the particle
energies on shell.  This formalism allows calculations of bound states
in three-dimensions (where much success has been shown) and provides
the formalism for embedding the result within a four-dimensional
covariant scattering theory.

In Sec.~\ref{sec:full-greens-function} we define the full three-body
Green's function.  Then, in Sec.~\ref{sec:3d-reduction} we examine the
three-dimensional reduction of the internal momenta of the Green's
function.  In Sec.~\ref{sec:3body-iterative-greens-function} we
examine the structure of the Green's function, organizing the
summation of graphs into pre-{} and post-factors, and a
three-dimensional iterative Green's function.  In
Sec.~\ref{sec:bound-state} we extract the bound-state equation from
the pole of the Green's function.  Finally, in
Sec.~\ref{sec:conclusion} we discuss our conclusions from this work.
We also provide three appendices: Appendix~\ref{sec:TOPT}, in which we
provide the rules for time-ordered perturbation theory for our model,
Appendix~\ref{sec:4->3}, in which we provide more details of the
reduction from four to three dimensions, and Appendix~\ref{sec:CS}, in
which we discuss cluster separability in our formalism.

\section{The full three-body Green's function}
\label{sec:full-greens-function}

The four-dimensional three-body Green's function is defined in
field theory as the six-point function
\begin{equation}
  {\cal G}(x_1,x_2,x_3;y_1,y_2,y_3)\equiv
  \langle 0 | \; 
  T[\psi_1(x_1)\psi_2(x_2)\psi_3(x_3)
  \bar\psi_3(y_3)\bar\psi_2(y_2)\bar\psi_1(y_1)]
  \; | 0 \rangle\;.
\label{eq:g-field-theory}
\end{equation}
Here we allow $\psi$ to represent either spin-1/2 or spin-0 particles,
and consider the three particles to be distinguishable.  Note that
with distinguishable particles, $\psi_1$ can only contract with
$\bar\psi_1$, and for spin-1/2 particles anticommutes with
$\bar\psi_2$ and $\bar\psi_3$.  The interaction is assumed to be a sum
of meson-nucleon interaction terms [see
Eq.~(\ref{eq:interaction-hamiltonian})].

\begin{figure}
    \centerline{ \epsfbox{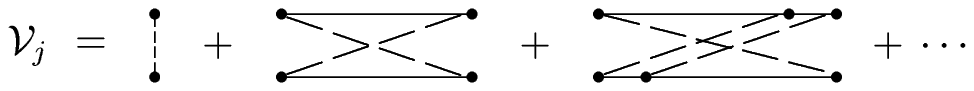} }
    \caption{The sum of two-body-irreducible graphs denoted by ${\cal
      V}_j$.  }
    \label{fig:two-body-irreducible}
  \end{figure}

The Bethe-Salpeter equation
\cite{salpeter:pr-84-1232,gell-mann:pr-84-350} for the three-body
Green's function may be derived from Eq.~(\ref{eq:g-field-theory}) by
expanding ${\cal G}$ in a perturbation series in the interaction
picture.   Rearranging the resulting Feynman graphs into two sets, the
two-body irreducible graphs and the iterative graphs, and denoting the
sum of two-body irreducible graphs for cluster $j$ as ${\cal V}_j$
(see Fig.~\ref{fig:two-body-irreducible}), 
%
%
we find
\begin{eqnarray}
  {\cal G} 
  &=& 
    -i g_1 g_2 g_3 
    - {\cal G} \sum_{j=1}^3 {\cal V}_j g_k g_l
  =
    {\cal G}_0 
    + {\cal G} \sum_j \left[ {\cal V}_j (ig_j)^{-1} \right] {\cal G}_0
\nonumber \\*
  &=& 
    {\cal G}_0 
    + {\cal G}_0 \sum_j \left[ {\cal V}_j (ig_j)^{-1} \right] {\cal G}
  \; ,
\label{eq:bse-4d}
\end{eqnarray}
where ${\cal G}_0 = -i g_1 g_2 g_3 $.  (For reviews of the
Bethe-Salpeter equation see Nakanishi \cite{nakanishi:sptp-43-1} and
Remiddi \cite{remiddi:ISPEF80-1}.)  Note that we are considering
distinguishable particles, and that the subscript of the
single-particle Green's function $g_j$ refers to the particle species.
For indistinguishable particles the products would need to be
symmetrized or antisymmetrized depending upon whether we are
considering spin-0 or spin-1/2 particles.  
The index $j$ of the interaction in Eq.~(\ref{eq:bse-4d}) conforms to
``odd particle out'' notation, such that ${\cal V}_j$ represents the
sum of two-body irreducible graphs between particles $k$ and $l$,
where ($jkl$) is an even permutation of ($123$).  The
``self-energy-summed'' single-particle propagator $g_j$ is
approximated by the free propagator using the physical mass,
\begin{equation}
  g_j(x',x) \approx 
  -i \langle 0 | \; T[ \psi_j(x') \bar \psi_j(x) ] \; | 0 \rangle 
  \; .
\label{eq:free-prop-approx}
\end{equation}
Although this analysis neglects Feynman diagrams representing
three-body interactions, we shall see later that Eq.~(\ref{eq:bse-4d})
contains {\em time-ordered} diagrams representing three-body
interactions.

\section{Three-dimensional reduction}
\label{sec:3d-reduction}

A three-dimensional reduction follows from decomposing graphs into
sums of distinct time intervals between consecutive meson emission and
absorption events, and then integrating over the duration of each
interval.

In order to establish some notation, the free-particle propagator is
written as the sum of a positive-{} and a negative-energy part
\begin{equation}
  g(p)=\frac{N^+({\bf p})}{p^0 - \epsilon({\bf p}) +i\eta}
      -\frac{N^-({\bf p})}{p^0 + \epsilon({\bf p}) -i\eta}\; ,
\end{equation}
where the on-shell energy of a particle is denoted by
\begin{equation}
  \epsilon_j ({\bf p}) \equiv \sqrt{ {\bf p}^2 + m_j^2 } \; ,
\end{equation}
and we define
\begin{equation}
  N^+({\bf p}) \equiv u({\bf p}) \bar u({\bf p})  \; ,
\end{equation}
and similarly
\begin{equation}
  N^-({\bf p}) \equiv - v(-{\bf p}) \bar v(-{\bf p}) \; .
\end{equation}
For spin-1/2 particles, $u({\bf p})$ is a positive-energy Dirac spinor,
$v(-{\bf p})$ is a negative-energy Dirac spinor, $\bar u({\bf p}) =
u^{\dag}({\bf p}) \gamma^0$,  and $\bar v(-{\bf p}) = v^{\dag}(-{\bf
p}) \gamma^0$.  Dirac spinors obey the Hermitian normalization
conditions: $u^{\dag}({\bf p}) u({\bf p}) = 1$, and $v ^{\dag}(-{\bf
p}) v(-{\bf p}) = 1$.   For spin-0 particles $u({\bf p}) = \bar u({\bf
p}) = 1/\sqrt{2 \epsilon}$, and $v(-{\bf p}) = - \bar v(-{\bf p}) =
1/\sqrt{2 \epsilon}$.  Combining these definitions with that of
$\Gamma$ provided in Eq.~(\ref{eq:gamma-sub-i}), we have $\Gamma
\bar{u}({\bf p}) u({\bf p}) = g_0 m/\epsilon$ and $\Gamma
\bar{v}(-{\bf p}) v(-{\bf p}) = - g_0 m/\epsilon$ for either spin-0 or
spin-1/2 particles.  This notation permits the analysis to proceed on a
common footing for both spins.

To facilitate the time-ordered analysis, we perform a Fourier
transformation of the time-like component of the momentum of $g(p)$,
\begin{eqnarray}
  \label{eq:g+-time}
    g(t',t;{\bf p}) 
  &\equiv& 
    \int \frac{dp^0}{(2\pi)} e^{-i p^0 (t' - t) } g(p)  
  \nonumber \\
  &=& 
    (-i) \theta(t' - t) N^+({\bf p}) e^{-i \epsilon({\bf p}) (t' - t)}
     -i  \theta(t - t') N^-({\bf p}) e^{ i \epsilon({\bf p}) (t' - t)}
  \; .
\end{eqnarray}
  
In order to simplify the analysis, two approximations are used:

(i) {\em Positive energy particles.\/}
    The contribution of negative-energy states to the propagator of
Eq.~(\ref{eq:g+-time}) is neglected.  This is the traditional
starting point in nuclear physics.  It breaks the covariance of the
theory but is generally believed to be commensurate with our present
understanding of nuclear forces.  Extension of the analysis to
incorporate negative-energy states will be left to the future.

(ii) {\em One boson exchange.\/}
  For NN interactions, a suitable effective interaction should
describe the NN phase shifts and deuteron binding.  This may be
accomplished by the use of a one-boson-exchange (OBE) interaction,
such that ${\cal V}_j$ is replaced by the first term in
Fig.~\ref{fig:two-body-irreducible}.

Consider an interaction Hamiltonian of the form,
\begin{equation}
  {\cal H}_{\text{I}} = 
  \left\{ 
    \begin{array}{l}
      \displaystyle
      \sum_{j=1}^3
      - g_0 : \bar \psi_j \varphi \psi_j:, 
      \quad \text{for spin-1/2,}\\
      \displaystyle
      \sum_{j=1}^3
      - 2 m_j g_0 : \Phi^{*}_j \varphi \Phi_j :, 
      \quad \text{for spin-0.}
    \end{array} 
  \right.
\label{eq:interaction-hamiltonian}
\end{equation}
where $g_0$ is a coupling constant.  Separate fields are introduced
for each particle in order to treat them as distinct particles.  

It is convenient to define a vertex factor as follows:  
  \begin{equation} 
    \Gamma_j \equiv 
    \left\{
      \begin{array}{l}
        g_0, \quad \text{for spin-1/2,}\\
        2 m_j g_0, \quad \text{for spin-0,}
      \end{array}
    \right.
  \label{eq:gamma-sub-i}
  \end{equation}
where the factor $2 m_j$ is introduced in order that both cases
have a common nonrelativistic limit.  Performing a Fourier transform
with respect to the timelike component of the momentum, and using the
approximation in which only the one-boson-exchange potential is
retained, produces
  \begin{equation}
    {\cal V}_j \simeq 
    - i \Gamma_k \Gamma_l \left\{
    (-i) \theta(t_k - t_l) \frac{e^{-i \omega (t_k - t_l)}}{2 \omega} 
    +
    (-i) \theta(t_l - t_k) \frac{e^{+i \omega (t_k - t_l)}}{2 \omega} 
    \right\} ,
  \label{eq:obe-time}
  \end{equation}
where $\omega = \sqrt{\mu^2 + {\bf q}^2}$, with $\mu$ being the meson
mass and ${\bf q}$ being its three momentum.  

The OBE interaction together with the use of the free-particle
propagator [Eq.~(\ref{eq:free-prop-approx})] means that the analysis
pertains to the ladder approximation Feynman graphs
\cite{nakanishi:sptp-43-1}.

Inserting Eqs.~(\ref{eq:g+-time}) and~(\ref{eq:obe-time}) into
Eq.~(\ref{eq:bse-4d}) leads to an equation for the Green's
function in which all of the ``internal'' integrations are over the
three momenta and the relative times.  Each contribution can be
decomposed into a sequence of time intervals between consecutive meson
emission and absorption events.  The integral over the time duration
of each such interval may be performed  analytically to produce the
time-ordered perturbation theory (TOPT) rules and corresponding graphs
(see Appendix \ref{sec:TOPT}).

In order to provide a few examples, consider subgraphs that contain no
initial or final particle lines, i.e., that are embedded within other
graphs such that interactions separate them from initial or final
particle lines.  
Time intervals in which three particles propagate freely correspond to 
\setlength{\unitlength}{1.666666pt}
\begin{equation}
    \setlength{\unitlength}{1.666666pt}
    \vcenter{
      \hbox{ 
        \begin{picture}(33,22)
  	\put(0,0){\line(1,0){33}}
  	\put(0,11){\line(1,0){33}}
  	\put(0,22){\line(1,0){33}}
        \end{picture}
      }
    }
    = \frac{1}{P^0 - \epsilon_1 - \epsilon_2 - \epsilon_3 +i\eta}
    \equiv G_0
\label{eq:G0}
\end{equation}
and time intervals beginning with emission of a meson and ending with
its absorption correspond to
\begin{equation}
    \setlength{\unitlength}{1.666666pt}
    \vcenter{
      \hbox{ 
        \begin{picture}(33,22)
  	\put(0,0){\line(1,0){33}}
  	\put(0,11){\line(1,0){33}}
  	\put(0,22){\line(1,0){33}}
  	\multiput(1,21.6666)(8,-2.6666){4}{\line(3,-1){6}}
  	\put(0,22){\circle*{2}}
  	\put(33,11){\circle*{2}}
        \end{picture}
      }
    }
    = 
    \frac{\Gamma_2 N_2^+}{\sqrt{2 \omega_3}}
    \frac{1}{P^0 - \epsilon_1 - \epsilon_2 - \epsilon_3 - \omega_3 
    +i\eta}
    \frac{\Gamma_1 N_1^+}{\sqrt{2 \omega_3}}
\label{eq:omega3}
\end{equation}
When a second meson is in flight during a time interval such as in
Eq.~(\ref{eq:omega3}), the contribution is modified to the form,
\begin{equation}
    \setlength{\unitlength}{1.666666pt}
    \vcenter{
      \hbox{ 
        \begin{picture}(66,28)(-16.5,-3)
  	\thicklines
  	\put(0,0){\line(1,0){33}}
  	\put(0,11){\line(1,0){33}}
  	\put(0,22){\line(1,0){33}}
  	\multiput(1,21.6666)(8,-2.6666){4}{\line(3,-1){6}}
  	\multiput(1,8)(8,-1.3333){4}{\line(6,-1){6}}
  	\put(0,22){\circle*{2}}
  	\put(33,11){\circle*{2}}
  	\put(-3,-3){\dashbox{1}(39,28){}}
  	\thinlines
  	\put(-16.5,0){\line(1,0){66}}
  	\put(-16.5,11){\line(1,0){66}}
  	\put(-16.5,22){\line(1,0){66}}
  	\multiput(-15.5,10.6666)(8,-1.3333){2}{\line(6,-1){6}}
  	\multiput(33,2.6666)(8,-1.3333){2}{\line(6,-1){6}}
  	\put(-16.5,11){\circle*{2}}
  	\put(49.5,0){\circle*{2}}
        \end{picture}
      }
    }
    \propto 
    \vcenter{
      \hbox{ 
        \begin{picture}(33,22)(0,0)
  	\put(0,0){\line(1,0){33}}
  	\put(0,11){\line(1,0){33}}
  	\put(0,22){\line(1,0){33}}
  	\multiput(1,21.6666)(8,-2.6666){4}{\line(3,-1){6}}
  	\multiput(1,8)(8,-1.3333){4}{\line(6,-1){6}}
  	\put(0,22){\circle*{2}}
  	\put(33,11){\circle*{2}}
        \end{picture}
      }
    }
    =
    \frac{\Gamma_2 N_2^+}{\sqrt{2 \omega_3}}
    \frac{1}{P^0 - \epsilon_1 - \epsilon_2 - \epsilon_3 - \omega_1 -
    \omega_3 + i\eta}
    \frac{\Gamma_1 N_1^+}{\sqrt{2 \omega_3}}
\label{eq:omega1-omega3}
\end{equation}
where labels of the exchanged bosons correspond to
``odd-particle-out'' notation. 
 In Eq.~(\ref{eq:omega1-omega3}), the
expression on the right-hand side corresponds to the time interval and
vertices within the dashed box on the left-hand side.  
For each distinct time interval there is a denominator
equal to the total energy minus the sum of the on-shell energies
of all of the particles present during that time interval. 
Although consideration was limited to the OBE interaction, the
time-ordered rules apply quite generally and, for example, the
``cross-box'' diagram shown yields:
  \setlength{\unitlength}{1.666666pt}
  \begin{eqnarray}
    \vcenter{
      \hbox{ 
        \begin{picture}(41.25,22)(0,0)
  	\put(0,0){\line(1,0){41.25}}
  	\put(0,11){\line(1,0){41.25}}
  	\put(0,22){\line(1,0){41.25}}
  	\multiput(1,21.6666)(8,-2.6666){4}{\line(3,-1){6}}
  	\put(0,22){\circle*{2}}
  	\put(33,11){\circle*{2}}
  	\multiput(9.25,11.3333)(8,2.6666){4}{\line(3,1){6}}
  	\put(8.25,11){\circle*{2}}
  	\put(41.25,22){\circle*{2}}
        \end{picture}
      }
    }
    &=&
    \frac{\Gamma_1 N_{1''}^+}{\sqrt{2 \omega'_3}}
    \frac{1}{P^0 - \epsilon''_1 - \epsilon_2 - \epsilon_3 - \omega'_3 +
      i\eta}
  \nonumber \\* && \times
    \frac{\Gamma_2 N_{2''}^+}{\sqrt{2 \omega_3}}
    \frac{1}{P^0 - \epsilon''_1 - \epsilon''_2 - \epsilon_3 - \omega_3
    - \omega'_3 +i\eta}
    \frac{\Gamma_2 N_{2'}^+}{\sqrt{2 \omega'_3}}
  \nonumber \\* && \times
    \frac{1}{P^0 - \epsilon''_1 - \epsilon'_2 - \epsilon_3 - \omega_3 +
      i\eta}
    \frac{\Gamma_1 N_{1'}^+}{\sqrt{2 \omega_3}}
  \end{eqnarray}
Note that there is an implied integration over the loop three momenta.
In general there are six time-ordered diagrams corresponding to
a single cross-box Feynman graph, differing by the time ordering of
vertices on one particle line with respect to those on the other
particle line.

\begin{figure}
    \centerline{ \epsfbox{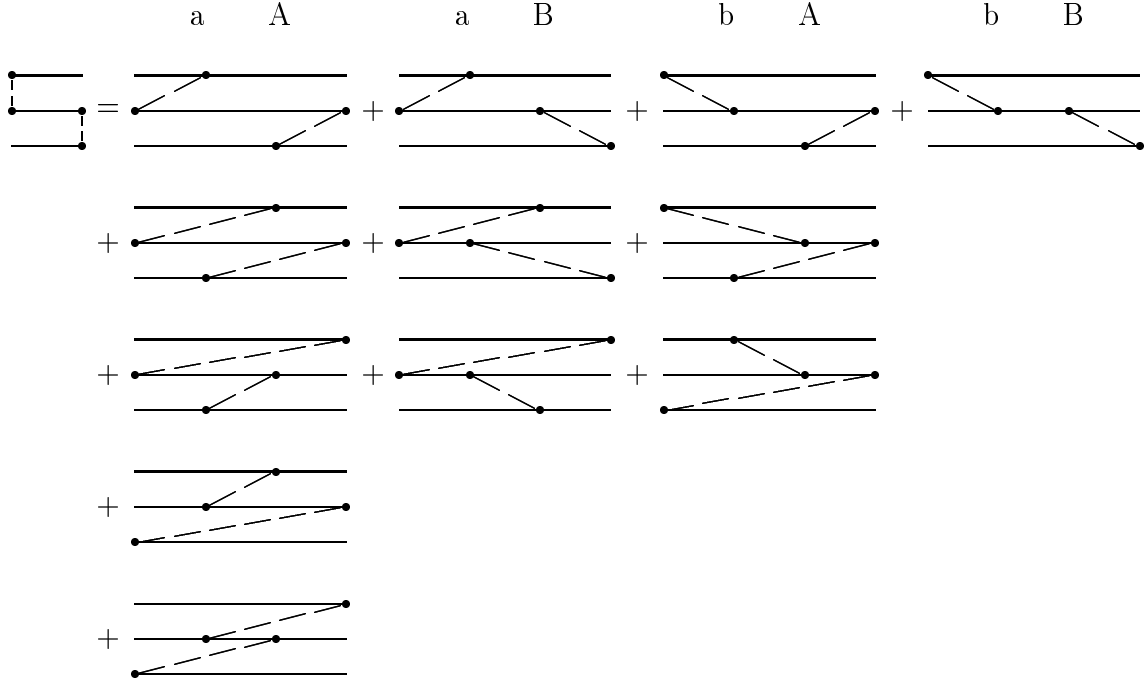} }
    \caption{By the reduction from four to three dimensions, the
      Feynman diagram on the left produces the 12 time-ordered
      diagrams on the right.  Note that four of the time-ordered
      diagrams are iterative like the original Feynman diagram, but
      that the other eight are not, and constitute a three-dimensional
      three-body force.}
    \label{fig:reclassify}
  \end{figure}

The transition from four to three dimensions necessitates a
reclassification of diagrams in terms of three-particle irreducibility
with respect to $G_0$ so as to distinguish between two-{} versus
three-body forces in the time-ordered formalism.  Consider the simple,
iterative Feynman (four-dimensional) diagram shown at the left of
Fig.~\ref{fig:reclassify}.
The reduction from four to three dimensions produces 12 time-ordered
diagrams, the sum of the first four being the iteration of the
two-body force.  The other eight, however, do not have the requisite
time interval with only the three particles propagating, and are hence
three-particle irreducible.  These diagrams, along with their
``sibling'' diagrams (where particle two interacts with particle three
before particle one), make up the three-body force $V_0^{31}$.  The
subscript denotes ``zero particles out'' while the superscript labels
the odd-particle-out exchanges of which this graph is composed.
Please see Appendix \ref{sec:4->3} for a fuller description of this
transition from four to three dimensions.

\section{The three-dimensional three-body Green's function}
\label{sec:3body-iterative-greens-function}

Having carried out the reduction to three dimensions and reclassified
the diagrams, we may re-sum the infinite series for ${\cal G}$.  
This sum may then be separated and factored.  First, we take cluster
separability into account, and then we factor the connected three-body
term into three pieces.

Cluster separability \cite{foldy:pr-122-275,wichmann:pr-132-2788}
means that a Green's function describing the propagation of clusters
of particles, when there are no interactions between the clusters,
consists of a product of independent factors, one for each cluster.
Each cluster's factor is the same as if the other clusters were not
present.
For three particles, we may have (a) no
particles interacting, (b) two particles interacting and the third a
spectator, and (c) all three particles interacting.  As these are the
only possible cases, cluster separability states that for the most
general Green's function we have
\begin{eqnarray}
  {\cal G}(p_{1f},p_{2f},p_{3f};p_{1i},p_{2i},p_{3i}) &=& 
  {\cal G}^{(1)}_{c;1}(p_{1f};p_{1i}) \,
  {\cal G}^{(1)}_{c;2}(p_{2f};p_{2i}) \,
  {\cal G}^{(1)}_{c;3}(p_{3f};p_{3i}) \,
\nonumber \\* &&
  {} + \sum_{j=1}^3 
    {\cal G}^{(1)}_{c;j}(p_{jf};p_{ji}) \,
    {\cal G}^{(2)}_{c;kl}(p_{kf},p_{lf};p_{ki},p_{li}) 
\nonumber \\* &&
  {} + {\cal G}^{(3)}_c(p_{1f},p_{2f},p_{3f};p_{1i},p_{2i},p_{3i})  \; .
\label{eq:cluster-separated-G}
\end{eqnarray}
We are denoting a fully connected $n$-particle Green's function for
particles $j,\ldots,k$ by ${\cal G}^{(n)}_{c;j,\ldots,k}$.  (We
omit the $1,2,3$ label for $n=3$.)    
Note that the fully connected part, ${\cal{G}}_c^{(3)}$, vanishes by
definition in the limit that one particle does not interact; the
correct cluster limit obtains from the disconnected diagrams in this
case.  Note also that if a three-body bound state exists, it
corresponds to a pole in ${\cal{G}}_c^{(3)}$ and has no contribution in
the other parts.  
Please see Appendix~\ref{sec:CS}
for a more complete discussion.

As we are primarily interested in obtaining the bound state, we
discard two sets of graphs: (a) disconnected graphs in which one or
more of the particles never interact, and (b) graphs in which there is
an initial or final particle in every three-body-reducible time
interval (i.e., in which there are no factors of $G_0$).  (See
Appendix~\ref{sec:CS}.)  After discarding these two sets of graphs, we
find that the sum of connected graphs in which there is at least one
fully internal time interval is factored into three parts,
\begin{equation}
  {\cal G}^{(3)}_{c} \longrightarrow 
  i  (2\pi)^4 \delta^{(4)}(P_f-P_i) 
  f_{\text{post}}^{(4\text{D}\rightarrow3\text{D})}
  G_3^{(3\text{D})}
  f_{\text{pre}}^{(3\text{D}\rightarrow4\text{D})}
  \; .
\label{eq:fG3f}
\end{equation}

We shall define $G_3$, $f_{\text{pre}}$, and $f_{\text{post}}$
shortly.  The superscripts $(4\text{D}\rightarrow3\text{D})$,
$(3\text{D})$, and $(3\text{D}\rightarrow4\text{D})$ label the
dependence of the three parts of the Green's function on the
initial-state and final-state relative energies.  The superscript
$(3\text{D})$ denotes that $G_3$ has no dependence upon the initial or
final relative energies.  As we have integrated out the internal
relative energies this function is purely three dimensional.  The
superscript $(4\text{D}\rightarrow3\text{D})$ on $f_{\text{post}}$
denotes that it depends upon the final relative energies of the
particles.  This function describes the transition from the
three-dimensional internal time intervals to the four dimensional
final state.  The function $f_{\text{pre}}$ performs a similar
function, depending upon the initial relative energies of the
particles.  As mentioned in Appendix~\ref{sec:CS}, this pattern of
energy dependence is present in any fully connected Green's function
that goes beyond the Born terms.  However, the functional forms of
the parts depend upon the number of particles involved.  

We define $G_3$ as
\begin{eqnarray}
  G_3 & \equiv & 
      G_0 \sum_{n=0}^\infty 
      \left[
	\left(\sum_{j=1}^3 V_j N^+_k N^+_l  + 
	V_0 N^+_1  N^+_2  N^+_3  \right)
	G_0\right]^n 
      \nonumber \\
  &=& G_0 + G_0 
      \left(\sum_{j=1}^3 V_j N^+_k N^+_l  + 
      V_0 N^+_1  N^+_2  N^+_3  \right)
      G_3 \; .
\label{eq:G3}
\end{eqnarray}
The three-body potential ($V_0 = V_0^{12} +  V_0^{23} + V_0^{31} +
V_0^{123} + \cdots$) is the sum of diagrams in which (a) all three
particles are interacting and (b) there are no
three-particle-reducible time intervals.  The three-dimensional three-body
Green's function $G_3$ represents the sum of the diagrams in which all
of the particles have interacted at least once, and will interact at
least once more.  This requirement ensures that interactions in $G_3$
are separated by the global propagator $G_0$ and do not contain any
dependence on the initial-state and final-state particle relative
energies, i.e., are fully three dimensional.  When there is a
three-body bound-state pole, it is contained in $G_3$.

Additional factors $f_{\text{pre}}$ and $f_{\text{post}}$ arise 
in Eq.~(\ref{eq:fG3f}) from sequences of two-body interactions
of particles $j$ and $k$ in the initial and final states, separated by
the two-body propagator for cluster $l$, $G_0^l \equiv
1/\left(P^0_{jk} - \epsilon_j -\epsilon_k +i\eta\right)$,
where $P^0_{jk} = p^0 _{j} + p^0 _k $ is the total energy of the pair. 
These factors may be expressed as
\begin{mathletters}
\label{eq:pre-and-post-factors}
\begin{eqnarray}
  f_{\text{post}} &=&
    \frac{N^+_1 N^+_2 N^+_3}{
      (p^0_{1f}-\epsilon_1 + i\eta)
      (p^0_{2f}-\epsilon_2 + i\eta) 
      (p^0_{3f}-\epsilon_3 + i\eta)} 
  \nonumber \\* && \times
    \left[
      \sum_{j=1}^3 
      \sum_{k\neq j} 
      \left( V_j N^+_k N^+_l G_0^j \right) \Omega^L_j
      \left( 
	V_k + \frac{V_0}{2} N^+_k
      \right) N^+_l N^+_j 
    \right]
    \; ,
  \\[\parsep]
  f_{\text{pre}} &=&
    \left[
      \sum_{j=1}^3 \sum_{k\neq j}
      \left(
	V_k + \frac{V_0}{2} N^+_k
      \right) 
      N^+_l N^+_j 
      \Omega^R_j \left( G_0^j V_j N^+_k N^+_l \right)
    \right]
  \nonumber \\* && \times
    \frac{1}{
      (p^0_{1i} - \epsilon_1 + i\eta)
      (p^0_{2i} - \epsilon_2 + i\eta)
      (p^0_{3i} - \epsilon_3 + i\eta)}
    \; ,
\end{eqnarray}
\end{mathletters}
where $\Omega^L_j$ and $\Omega^R_j$ are the left and right two-body
wave operators for cluster $j$, defined as
\begin{mathletters}
\begin{eqnarray}
  \Omega^L_j &\equiv& 
  \sum_{m=0}^\infty 
  \left(
    V_j N^+_k N^+_l 
    \frac{1}{p^0_{k}+p^0_{l}-\epsilon_k-\epsilon_l +i\eta}
  \right)^m
  = \sum_{m=0}^\infty 
  \left(
  V_j N^+_k N^+_l G_0^j
  \right)^m
  \; ,
\\
  \Omega^R_j &\equiv& 
  \sum_{m=0}^\infty 
  \left(
    \frac{1}{p^0_{k}+p^0_{l}-\epsilon_k-\epsilon_l +i\eta}
    V_j N^+_k N^+_l 
  \right)^m
  \; .
\end{eqnarray}
\end{mathletters}
These wave operators transform the two-body free propagator
$G_0^j$ into the full two-body Green's function $G_2^j$:
\begin{eqnarray}
  G_2^j&=&G_0^j \Omega^L_j= \Omega^R_j G_0^j
  \nonumber \\ 
  &=& G_0^j  + G_0^j V_j N^+_k N^+_l G_2^j \; .
\label{eq:twobody-G}
\end{eqnarray}
Through these wave operators, the pre-{} and post-factors
($f_{\text{pre}}$ and $f_{\text{post}}$) contain the two-body
bound-state poles for the different possible clusterings of particles
within the initial and final states.  Note that when all particles in
the initial and final states are on the mass shell, $P^0 _{jk} = P^0 -
\epsilon _l$, and $G^l _0 = G_0 $.   For consideration of
interactions of the bound state with, say, a photon, one needs the
full structure of ${\cal G}$, including the pre- and post-factors that
allow for breakup.  However, the three-body bound state is determined
from consideration of $G_3$ as defined by Eq.~(\ref{eq:G3}).  

\section{Bound-state equation }
\label{sec:bound-state}

Assume that there is a pole in $G_3$ at $P^0=E_{\text{B}}({\bf P})$,
where $E_{\text{B}}({\bf P}) = \sqrt{M_{\text{B}}^2 + {\bf P}^2 }$,
and $M_{\text{B}}$ is the bound-state mass.  
To find the bound-state equation, we write $G_3$ as
\begin{equation}
  G_3 = \frac{\left| \psi \right\rangle 
  \left\langle \psi \right|}
  {P^0-E_{\text{B}}+i\eta} + R
  \; ,
\label{eq:g-poles}
\end{equation}
and therefore
\begin{equation}
  {\cal G}^{(3)}_{c} = 
  i  (2\pi)^4 \delta^{(4)}(P_f-P_i) 
  \left[
    \frac{
      f_{\text{post}} 
      \left| \psi \right\rangle
      \left\langle \psi \right|
      f_{\text{pre}} 
    } {P^0-E_{\text{B}}+i\eta}
    +
    f_{\text{post}} R f_{\text{pre}}
  \right] \; ,
\end{equation}
where $\psi$ is the three-dimensional TOPT analog of the
Bethe-Salpeter wave function for the bound state, and $R$ is regular
at the bound-state pole.  Inserting Eq.~(\ref{eq:g-poles}) into the
second line of Eq.~(\ref{eq:G3}), taking the residue at $P^0
\rightarrow E_{\text{B}}$, and rearranging, we have
\begin{eqnarray}
  \Psi^{E_{\text{B}}}({\bf p}_1,{\bf p}_2,{\bf p}_3) &=& 
  G_0^{E_{\text{B}}}({\bf p}_1,{\bf p}_2,{\bf p}_3)
  \int \frac{d{\bf p}'_1}{(2\pi)^3} \frac{d{\bf p}'_2}{(2\pi)^3}
    \frac{d{\bf p}'_3}{(2\pi)^3}
\nonumber \\ && \times
  \left[
    \sum_{j=1}^3 
    \widetilde{V}_j^{E_{\text{B}}}
      ({\bf p}_k, {\bf p}_l, {\bf p}_j ; 
      {\bf p}'_k, {\bf p}'_l, {\bf p}_j)
    (2\pi)^3 \delta^{(3)}({\bf p}_j - {\bf p}'_j) \right.
\nonumber \\ && \phantom{\times\left[\vphantom{\sum_{j=1}^3} \right.} 
    \left. \vphantom{\sum_{j=1}^3}
    +
    \widetilde{V}_0^{E_{\text{B}}}
      ({\bf p}_1,{\bf p}_2,{\bf p}_3;
      {\bf p}'_1,{\bf p}'_2,{\bf p}'_3)
  \right]
  \Psi^{E_{\text{B}}}({\bf p}'_1,{\bf p}'_2,{\bf p}'_3)
  \; ,
\label{eq:3B-bound-state}
\end{eqnarray}
where we have defined 
\begin{equation}
  \Psi^{E_{\text{B}}}({\bf p}_1,{\bf p}_2,{\bf p}_3) \equiv  
  \bar u_1({\bf p}_1) \bar u_2({\bf p}_2) \bar u_3({\bf p}_3) \;
  \psi^{E_{\text{B}}}({\bf p}_1,{\bf p}_2,{\bf p}_3) \; ,
\end{equation}
and
\begin{mathletters}
\label{eq:new-potential}
\begin{equation}
  \widetilde{V}_j^{E_{\text{B}}} \equiv
  \left\{
    \begin{array}{l}
      \bar u_k({\bf p}_k) \bar u_l({\bf p}_l) \; 
      \left[ 
      V_j^{E_{\text{B}}}
        ({\bf p}_k, {\bf p}_l, {\bf p}_j ; 
        {\bf p}'_k, {\bf p}'_l, {\bf p}_j) \;
      \right] 
      u_k({\bf p}'_k) u_l({\bf p}'_l), 
      \quad \text{spin-1/2;}
    \\[1.5\parsep]
    \displaystyle
      \sqrt{ \frac{m_k m_l}
        { \epsilon_k({\bf p}_k) \epsilon_l({\bf p}_l)} } \; 
      \left[ 
      V_j^{E_{\text{B}}}
        ({\bf p}_k, {\bf p}_l, {\bf p}_j ; 
        {\bf p}'_k, {\bf p}'_l, {\bf p}_j) \;
      \right] 
      \sqrt{ \frac{m_k m_l}
        { \epsilon_k({\bf p}'_k) \epsilon_l({\bf p}'_l)} }, 
      \quad \text{spin-0.}
    \end{array}
  \right.
\end{equation}
\begin{equation}
  \widetilde{V}_0^{E_{\text{B}}} \equiv
  \left\{
    \begin{array}{l}
      \bar u_1({\bf p}_1) \bar u_2({\bf p}_2) \bar u_3({\bf p}_3) 
      \left[ 
      V_0^{E_{\text{B}}}
        ({\bf p}_1,{\bf p}_2,{\bf p}_3;
        {\bf p}'_1,{\bf p}'_2,{\bf p}'_3)
      \right] 
      u_1({\bf p}'_1)  u_2({\bf p}'_2) u_3({\bf p}'_3), 
    \quad \text{spin-1/2;}
    \\[1.5\parsep]
    \displaystyle
      \sqrt{ \frac{m_1 m_2 m_3}
        {\epsilon_1({\bf p}_1) 
        \epsilon_2({\bf p}_2) 
        \epsilon_3({\bf p}_3)} } \;
      \left[ 
      V_0^{E_{\text{B}}}
        ({\bf p}_1,{\bf p}_2,{\bf p}_3;
        {\bf p}'_1,{\bf p}'_2,{\bf p}'_3) \;
      \right] 
      \sqrt{ \frac{m_1 m_2 m_3}
        {\epsilon_1({\bf p}'_1) 
        \epsilon_2({\bf p}'_2) 
        \epsilon_3({\bf p}'_3)} }, 
    \quad \text{spin-0.}
    \end{array}
  \right.
\end{equation}
\end{mathletters}
Here the two-body interaction $\widetilde{V}_j^{E_{\text{B}}}$, which
consists of a sum of two-particle irreducible time-ordered graphs
between particles $k$ and $l$ ($V_j^{E_{\text{B}}}$), multiplied by
spinor factors, depends upon the momentum of the noninteracting
particle ${\bf p}_j$.  This is due to the term $P^0 - \epsilon_j({\bf
p}_j)$ in the denominator of the potential [see, for example,
Eq.~(\ref{eq:omega3})].  As noted earlier, the connected part
${\cal{G}}_c^{(3)}$ vanishes by definition if there is a
noninteracting particle.  Therefore, cluster separability is
unaffected by dependence of the two-body interactions that are
internal to ${\cal{G}}_c^{(3)}$ on the momentum of the spectator.  The
bound state for a two-body cluster in the limit that the third
particle does not interact (i.e., for sufficiently short-range
interactions and when there are no zero-energy bound states) derives
from  ${\cal G}^{(2)}_c$  and the potentials internal to it have no
dependence on the spectator momentum.

Equation~(\ref{eq:3B-bound-state}) is a Schr{\"o}dinger-like
relativistic equation: it is three dimensional, has a global
relativistic propagator [Eq.~(\ref{eq:G0})], and reduces to the
Schr{\"o}dinger equation in the nonrelativistic limit.  Also note that
our interaction has energy dependence and hence retardation.

\section{Conclusion}
\label{sec:conclusion}

We have examined the full three-body Green's function for the case of
one-boson-exchange interactions and positive-energy spin-0 or spin-1/2
particles.  We expanded out the Bethe-Salpeter equation for the
Green's function into an infinite series of four-dimensional graphs.
After performing a Fourier transformation of the internal energies
into relative times, we integrated over the relative times, leaving
an expansion of the full Green's function which only involved
three-dimensional internal variables (and hence integrations), while
still depending upon the full four-dimensional nature of the initial
and final states.  

Concentrating upon those graphs which contribute to the three-body
bound-state pole, we re-summed the series into three factors: a
three-dimensional Green's function $G_3$ obeying an iterative
equation, and pre-{} and post-factors which link the three-dimensional
$G_3$ to the off-shell states of the four-dimensional theory.  The
bound-state equation was then extracted from $G_3$ and shown to have a
Schr{\"o}dinger-like structure involving a global relativistic
propagator.  The one-boson-exchange potential was shown to be
augmented by factors which for spin-1/2 particles are plane-wave spinors,
and for spin-0 particles are kinematical factors. Although the
bound state is determined without reference to them, the pre- and
post-factors are needed when interactions are considered.  The current
must include off-shell factors to account for the introduction of
four-momentum into the graph through the interaction.  The complete
analysis of these currents, however, is left to future papers.

Numerical calculations involving three bosons are under way.  They are
based upon Eq.~(\ref{eq:3B-bound-state}) in the limit in which
$\widetilde{V}_0 \rightarrow 0$, and compare the full (retarded)
$\widetilde{V}_j$ to an instant approximation, as well as comparing
these forms to those proposed by others.  Full relativistic kinematics
are used in conjunction with these relativistic interactions.

\acknowledgments
P.C.D. thanks the Nuclear Physics Group at the University of New
Hampshire for use of their office space and their kind hospitality.
Support for this work by the U.S. Department of Energy under Grant
No.\ DE-FG02-93ER-40762 is gratefully acknowledged.

                              \appendix
\section{Time-ordered perturbation-theory rules}
\label{sec:TOPT}

The rules for calculating graphs in a time-ordered perturbation theory
using positive-energy particles are

\begin{enumerate}
  \item Assign an overall factor of
        \begin{itemize}
          \item[(i)]   $ -i $ if no particles interact,
          \item[(ii)]  $ \phantom{-{}}1 $ if only two particles interact,
          \item[(iii)] $ \phantom{-{}}i $ if all three particles interact.
        \end{itemize}
  \item Assign a factor of 
            $(2 \pi)^4 \, 
            \delta^{(4)}\!\left[\sum\left( p_f -  p_i \right)\right]$,
        where the sum is
        over interacting particles. 
  \item To each particle line with no interactions between the initial
        and final states, assign a factor of
        \[ 
          (2 \pi)^4 \, 
          \delta^{(4)}\!\left(p_f - p_i\right) 
          N^{+}({\bf p}_i)  
          \; .
        \]
  \item To each final particle line emerging from it's last
        interaction, assign a factor of
        \[ 
          \frac{N^{+}({\bf p}_f)}{p_f^0 - \epsilon + i\eta} \; . 
        \]
  \item For each vertex on particle $j$, 
    \begin{itemize}
    \item[(i)]
      Conserve 3-momentum, 
    \item[(ii)]
      assign a factor of:
      \[ \frac{\Gamma_j N_j^+({\bf p}'_j)}
	{\sqrt{2 \omega( \left|{\bf p}_j - {\bf p}'_j \right| ) } } 
        \; , 
      \]
      where ${\bf p}'_j$ is associated with an earlier time than 
      that with which ${\bf p}_j$ is associated.
    \end{itemize}
  \item For each unconstrained 3-momentum ${\bf p}$, assign a factor
    of
      \[ 
        \int \frac{d{\bf p}}{(2\pi)^3} \; . 
      \]
  \item To each time slice between vertices, assign a factor of:
     \[ 
       \frac{1}{ P^0 - E_1 - E_2 - E_3 - \sum_m \omega_m + i\eta} \; , 
     \]
     where,
       \[ 
         E_n\equiv 
         \left\{
           \begin{array}{l}
             \epsilon_n=\sqrt{{\bf p}^2_n+m^2_n},
             \quad \text{for internal particles;} 
           \\
             p^0_n, 
             \quad \text{for external particles,}
           \end{array}
         \right.
	\]
     and $m$ ranges over the exchanged bosons existing during the time
     slice.
     \label{energy-rules:interaction-factor}
    \item
      For each initial line, a factor of
      \[ 
        \frac{1}{p_{i}^0 - \epsilon + i\eta} \; .
      \]
\end{enumerate}

\section{Reduction from four to three dimensions}
\label{sec:4->3} 

\begin{figure}
    \centerline{ \epsfbox{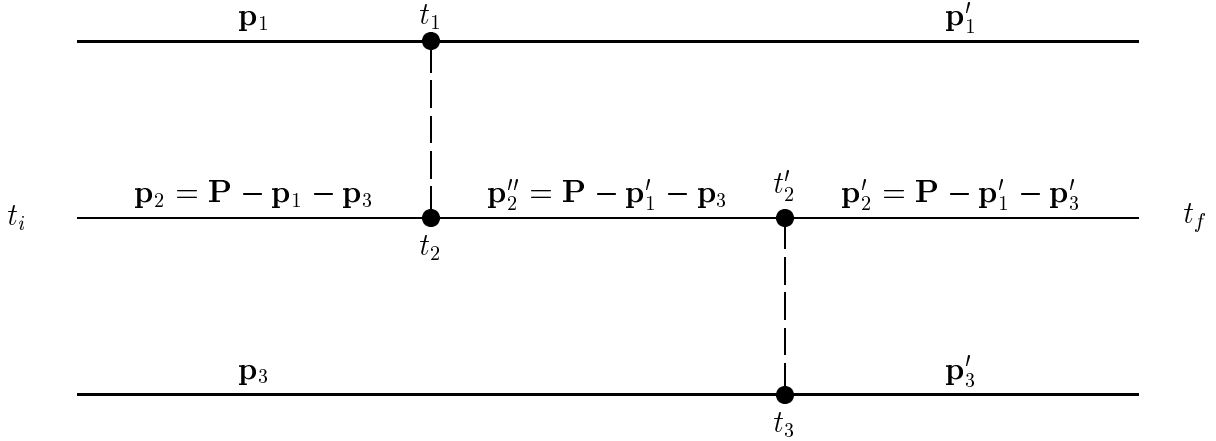} }
    \caption{An enlargement of the Feynman diagram on the left of
      Fig.~\protect{\ref{fig:reclassify}}.  Here we also add the time
      and three-momentum labels used in
      expression~(\protect{\ref{eq:one-twelve}}) to the particle
      lines.}
    \label{fig:feynman}
  \end{figure}

In Fig.~\ref{fig:reclassify} we have the Feynman diagram representing
an interaction between particles one and two, ``followed'' by an
interaction between particles two and three.   As in
Eqs.~(\ref{eq:G0}-\ref{eq:omega1-omega3}), we assume that this graph
is embedded within other graphs.  Using the notation of
Fig.~\ref{fig:feynman}, our positive-energy approximation, and
performing a Fourier transformation of the time-component of momentum,
we have
%
%
\begin{eqnarray}
  &&\int dt_f 
    \left\{ e^{iP^0_f t_f}  
    \;(-i)\theta(t_f-t_1) e^{-i\epsilon'_1(t_f-t_1)}
    \;(-i)\theta(t_f-t'_2) e^{-i\epsilon'_2(t_f-t'_2)}
    \;(-i)\theta(t_f-t_3) e^{-i\epsilon'_3(t_f-t_3)}
    \right\}
\nonumber \\
  && \times
  \int dt_i (-1) e^{-iP^0_i t_i}
  \int dt_1 dt_2 dt'_2 dt_3
\nonumber \\
  && \phantom{\quad} \times
  \left( 
  \frac{i\Gamma_2 \; i\Gamma_3}{2 \omega_1({\bf p}_3 - {\bf p}'_3)}
  \right)
  i \left[
  (-i)\overbrace{\theta(t'_2-t_3)}^{\text{A}}e^{-i\omega_1(t'_2-t_3)} + 
  (-i)\overbrace{\theta(t_3-t'_2)}^{\text{B}}e^{+i\omega_1(t'_2-t_3)}
  \right]
\nonumber \\
  &&\phantom{\qquad2\omega_1({\bf p}_3 - {\bf p}'_3)\times}\times
  i \left[ (-i)\theta(t'_2-t_2)N_{2''}^+ e^{-i\epsilon''_2(t'_2-t_2)} 
    \right] 
  i \left[ (-i)\theta(t_3-t_i)N_{3}^+ e^{-i\epsilon_3(t_3-t_i)} 
    \right] 
\label{eq:one-twelve}
\\ &&\phantom{\quad}\times
  \left(
  \frac{ i\Gamma_1 \; i\Gamma_2}{2 \omega_3({\bf p}_1 - {\bf p}'_1)}
  \right)
  i \left[
  (-i)\overbrace{\theta(t_1-t_2)}^{\text{a}}e^{-i\omega_3(t_1-t_2)} 
  +
  (-i)\overbrace{\theta(t_2-t_1)}^{\text{b}}e^{+i\omega_3(t_1-t_2)}
  \right]
\nonumber \\
  &&\phantom{\qquad2\omega_1({\bf p}_3 - {\bf p}'_3)\times}\times
  i \left[ (-i)\theta(t_1-t_i)N_{1}^+ e^{-i\epsilon_1(t_1-t_i)} 
    \right] 
  i \left[ (-i)\theta(t_2-t_i)N_{2}^+ e^{-i\epsilon_2(t_2-t_i)} 
    \right] 
\nonumber
\end{eqnarray}
where the first two lines of expression~(\ref{eq:one-twelve}) account
for the fact that this graph is embedded within another graph,
and we have labeled the exchanged-boson $\theta$~functions A, B, a,
and~b.  Using the identity
\begin{equation}
  \theta(x-y) \theta(x'-y) = 
  \theta(x-x') \theta(x'-y) + \theta(x'-x) \theta(x-y)
  \; ,
\end{equation}
one can expand the expression (\ref{eq:one-twelve}) into the 12 terms
shown in Fig.~\ref{fig:reclassify}.  The labels above the columns show
from which $\theta$ functions each column originates.  The top four
diagrams are three-particle reducible, and are therefore representable
as iterations of the two-body force.  The other eight diagrams are
part of the three-dimensional three-body force.

Let us choose the third diagram in the (aA) column of
Fig.~\ref{fig:reclassify}.  This diagram results from the $\theta$
functions 
\begin{equation}
  \theta(t_f - t_1) \theta(t_1 - t'_2) \theta(t'_2 - t_3) 
  \theta(t_3 - t_2) \theta(t_2 - t_i) \; .
\end{equation}
Defining the time interval variables
\begin{equation}
  \tau_5 = t_f - t_1,  \quad  \tau_4 = t_1 - t'_2,  \quad  
  \tau_3 = t'_2 - t_3,  \quad  \tau_2 = t_3 - t_2, \quad  
  \tau_1 = t_2 - t_i, \quad  \tau_0 = t_i \; ,
\end{equation}
and noting that $|\partial(t_f, t_1, t'_2, t_3, t_2,
t_i)/\partial(\tau_5, \tau_4, \tau_3, \tau_2, \tau_1, \tau_0)| = 1$,
Eq.~(\ref{eq:one-twelve}) has the form (for this particular
combination of $\theta$ functions)
\begin{eqnarray}
  &&\int d\tau_5 d\tau_4 d\tau_3 d\tau_2 d\tau_1 d\tau_0 \;
    \theta(\tau_5) \theta(\tau_4) \theta(\tau_3) 
    \theta(\tau_2) \theta(\tau_1) \theta(\tau_0)
\nonumber \\
  && \quad \times
    (-i)
    e^{i P^0_f ( \tau_5 + \tau_4 + \tau_3 + \tau_2 + \tau_1 ) }
    e^{i\tau_0 ( P^0_f - P^0_i) }
    \; e^{-i\epsilon'_1(\tau_5)}
    \; e^{-i\epsilon'_2(\tau_5 + \tau_4)}
    \; e^{-i\epsilon'_3(\tau_5 + \tau_4 + \tau_3)}
\nonumber \\
  && \quad \times
  \left( 
    \frac{\Gamma_2 \; \Gamma_3}{2 \omega_1({\bf p}_3 - {\bf p}'_3)}
  \right)
   \left[           e^{-i\omega_1    (\tau_3)} \right]
   \left[ N_{2''}^+ e^{-i\epsilon''_2(\tau_3 + \tau_2)} \right] 
   \left[ N_{3}^+  e^{-i\epsilon_3 (\tau_2 + \tau_1)} \right] 
\label{eq:aA3} \\ 
  && \quad \times
  \left(
    \frac{ \Gamma_1 \; \Gamma_2}{2 \omega_3({\bf p}_1 - {\bf p}'_1)}
  \right)
  \left[          e^{-i\omega_3   (\tau_4 + \tau_3 + \tau_2)} \right]
  \left[ N_{1}^+ e^{-i\epsilon_1(\tau_4 + \tau_3 + \tau_2 + \tau_1)}
  \right] 
  \left[ N_{2}^+ e^{-i\epsilon_2(\tau_1)} \right] \; .
\nonumber
\end{eqnarray}
Using the identities
\begin{mathletters}
\begin{equation}
  \int_{-\infty}^{\infty} d\tau \; e^{i\tau ( P^0_f - P^0_i) }
  = 2 \pi \; \delta( P^0_f - P^0_i)
  \; ,
\end{equation}
\begin{equation}
  \int_{-\infty}^{\infty} d\tau \; 
  \theta(\tau) \; e^{i\tau (a - b + i\eta) }
  = \frac{i}{a - b + i\eta}
  \; ,
\end{equation}
\end{mathletters}
we can calculate the integrals in Eq.~(\ref{eq:aA3}).  Simplifying the
result, we have
\begin{eqnarray}
  &&
    2 \pi \; \delta( P^0_f - P^0_i)
    \frac{1}{P^0_f -\epsilon'_1 -\epsilon'_2 -\epsilon'_3 +i\eta}
  \nonumber \\* && \times
    \left[
    \frac{ \Gamma_1 \; N_{1}^+}{\sqrt{2 \omega_3({\bf p}_1 - {\bf
    p}'_1)}} \;
    \frac{1}
      {P^0_f -\epsilon_1 -\epsilon'_2 -\epsilon'_3 -\omega_3 +i\eta}
    \right.
  \nonumber \\* && \quad \times
    \frac{ \Gamma_2 \; N_{2''}^+ }{\sqrt{2 \omega_1({\bf p}_3 - {\bf
    p}'_3)}} \;
    \frac{1}
      {P^0_f -\epsilon_1 -\epsilon''_2 -\epsilon'_3 -\omega_1
      -\omega_3 + i\eta} \;
    \frac{ \Gamma_3 \; N_{3}^+}
      {\sqrt{2 \omega_1({\bf p}_3 - {\bf p}'_3)}}
  \nonumber \\* && \times
    \left.
    \frac{1}
      {P^0_f -\epsilon_1 - \epsilon''_2 -\epsilon_3 -\omega_3
      +i\eta} \;
    \frac{ \Gamma_2 \; N_{2}^+}{\sqrt{2 \omega_3({\bf p}_1 - {\bf
    p}'_1)}} \right]
    \frac{1}
      {P^0_f -\epsilon_1 -\epsilon_2 -\epsilon_3 +i\eta} \; .
\label{eq:V0-aA3}
\end{eqnarray}

This result agrees with the rules given in Appendix.~\ref{sec:TOPT}.
The effect of the pre- and post-factors on this ``internal'' graph is
to introduce the total energy for the initial and final states.  Each
time interval corresponds to a denominator with the on-shell energy of
each existing particle subtracted from the total energy.  The graph
begins and ends with a $G_0$ factor.  Note that the graph in
Fig.~(\ref{fig:reclassify}) does not include these $G_0$ factors.
Instead it includes only those factors within the brackets.  These
factors are also the ones which contribute to $V_0$.

\section{Cluster separability and TOPT}
\label{sec:CS}

Cluster separability (CS) implies that if we describe particles
propagating using a Green's function, and if one cluster of particles
does not interact with another cluster of particles, then we can perform
a separation of variables (between these two clusters) on the Green's
function \cite{foldy:pr-122-275,wichmann:pr-132-2788}.

Let us describe three distinguishable particles propagating with
initial momenta $p_1,p_2,p_3$ and final momenta $p'_1,p'_2,p'_3$ by
the Green's function ${\cal G}(p'_1,p'_2,p'_3;p_1,p_2,p_3)$.  CS tells
us that in the absence of interactions, we must have
\begin{equation}
  {\cal G}(p'_1,p'_2,p'_3;p_1,p_2,p_3) \longrightarrow
  {\cal G}^{(1)}_{c;1}(p'_1;p_1) 
  {\cal G}^{(1)}_{c;2}(p'_2;p_2) 
  {\cal G}^{(1)}_{c;3}(p'_3;p_3) \; .
\end{equation}
If we have only two of the particles interacting (say 1 and 2),
we have
\begin{equation}
  {\cal G}(p'_1,p'_2,p'_3;p_1,p_2,p_3) \longrightarrow
  {\cal G}^{(2)}_{c;12}(p'_1,p'_2;p_1,p_2) 
  {\cal G}^{(1)}_{c;3}(p'_3;p_3) \; .
\end{equation}
Finally, if all three particles interact, we have the fully
connected Green's function
\begin{equation}
  {\cal G}(p'_1,p'_2,p'_3;p_1,p_2,p_3) \longrightarrow
  {\cal G}^{(3)}_c(p'_1,p'_2,p'_3;p_1,p_2,p_3)  \; .
\end{equation}
As these are the only possible cases, CS states that for the most
general three-body Green's function we have
Eq.~(\ref{eq:cluster-separated-G}), which we restate here
\begin{eqnarray}
  {\cal G}(p'_1,p'_2,p'_3;p_1,p_2,p_3) &=& 
  {\cal G}^{(1)}_{c;1}(p'_1;p_1) 
  {\cal G}^{(1)}_{c;2}(p'_2;p_2) 
  {\cal G}^{(1)}_{c;3}(p'_3;p_3)
\nonumber \\* &&
  {} + \sum_{i=1}^3 {\cal G}^{(2)}_{c;jk}(p'_j,p'_k;p_j,p_k) 
    {\cal G}^{(1)}_{c;i}(p'_i;p_i)
\nonumber \\* &&
  {} + {\cal G}^{(3)}_c(p'_1,p'_2,p'_3;p_1,p_2,p_3)  \; .
  \eqnum{\ref{eq:cluster-separated-G}}
\end{eqnarray}

We are working in momentum space, where the Green's function is simply
the result of performing a Fourier transform on the configuration
space Green's function.  For the $n$-body system, 
\begin{equation}
  {\cal G}(p'_1,\ldots,p'_n;p_1,\ldots,p_n) \equiv 
  \int \prod_{i=1}^{n} d^4 x_i\,d^4 y_i\, 
  e^{i (p'_i x_i - p_i y_i)}
  {\cal G}(x_1,\ldots,x_n;y_1,\ldots,y_n)  \; .
\label{eq:FT-of-G}
\end{equation}
Please note that some authors 
choose to note the Dirac $\delta$ functions explicitly, by factoring them
out of the momentum space Green's functions.  In this case, the left
hand side of Eq.~(\ref{eq:FT-of-G}) would be 
\[
  (2\pi)^4 \delta^{(4)}(p'_1+\cdots+p'_n-p_1-\cdots-p_n) 
  {\cal G}(p'_1,\dots,p'_n;p_1,\ldots,p_n) \equiv \cdots \; .
\]
We have chosen not to perform this separation.

\subsection{The TOPT case}
Let us now examine the results of CS in the context of the three-body
TOPT rules given in Appendix~\ref{sec:TOPT}.

\subsubsection{No interactions}

This case is the simplest, giving
\begin{eqnarray}
  {\cal G}(p'_1,p'_2,p'_3;p_1,p_2,p_3) &=& 
  (-i) 
  \frac{(2\pi)^4 \delta^{(4)} (p_1-p'_1) N^+({\bf p}_1)}
    {p^0_1 - \epsilon_1 + i \eta}
  \nonumber \\* &&
  \times
  \frac{(2\pi)^4 \delta^{(4)} (p_2-p'_2) N^+({\bf p}_2)}
    {p^0_2 - \epsilon_2 + i \eta}
  \nonumber \\* &&
  \times
  \frac{(2\pi)^4 \delta^{(4)} (p_3-p'_3) N^+({\bf p}_3)}
    {p^0_3 - \epsilon_3 + i \eta}
  \; ,
\end{eqnarray}
which implies that
\begin{equation}
  {\cal G}^{(1)}_{c;1}(p'_1;p_1) = 
  i g_1(p'_1;p_1) = 
  i
  (2\pi)^4 \delta^{(4)} (p_1-p'_1) 
  \frac{N^+({\bf p}_1)}{p^0_1 - \epsilon_1 + i \eta}
  \; .
\end{equation}
%

\subsubsection{Two particles interacting}

For definiteness we will assume that particles 1 and 2 are
interacting, and that particle 3 is the spectator.  This implies that
of the four interactions $V_0$, $V_1$, $V_2$, and $V_3$, only $V_3$ is
nonzero.  

From the TOPT rules given in Appendix~\ref{sec:TOPT}, we obtain the
Green's function ${\cal G}$ for the case of particles 1 and 2
interacting.  It is possible to write it in operator notation (defined
shortly) in the form
\begin{eqnarray}
    \lefteqn{
      {\cal G}(p'_1,p'_2,p'_3;p_1,p_2,p_3)
    }
  \nonumber \\* &\quad=&
    \Biggl\{
      i
      (2\pi)^4 \delta^{(4)} (p_3 - p'_3) 
      \frac{N^+_3} {p^0_3 - \epsilon_3 + i \eta}
    \Biggr\}
  \nonumber \\* && \times
    \Biggl\{
      (-i)
      (2\pi)^4 \delta^{(4)} (p_1 + p_2 - p'_1 - p'_2) 
      \frac{N^+_1} {p'^0_1 - \epsilon'_1 + i \eta} \,
      \frac{N^+_2} {p'^0_2 - \epsilon'_2 + i \eta}
  \nonumber \\* && \phantom{\times} \times
      \Biggl(
        \left[
          V_3^{(4\text{D})} N^+_1 N^+_2 
        \right]
        +
        \left[
          V_3^{(4\text{D}\rightarrow3\text{D})} N^+_1 N^+_2  
          \left(
            G_0^3 \Omega^L_3
          \right)
          V_3^{(3\text{D}\rightarrow4\text{D})} N^+_1 N^+_2
        \right]
      \Biggr)
  \nonumber \\* && \phantom{\times} \times
      \frac{1} {p^0_1 - \epsilon_1 + i \eta} \,
      \frac{1} {p^0_2 - \epsilon_2 + i \eta}
    \Biggr\}
\label{eq:G12-omega}
  \\ &\quad=&
      {\cal G}^{(1)}_{c;3}(p'_3;p_3) \,
      {\cal G}^{(2)}_{c;12}(p'_1,p'_2;p_1,p_2)
\end{eqnarray}
where the last line identifies ${\cal
G}^{(2)}_{c;12}(p'_1,p'_2;p_1,p_2)$.
First we will define the different forms of the potential $V_3$, and
then we will define the global propagator for cluster $l$, $G_0^l$,
and the wave operator $\Omega_3$.

As shown in item~\ref{energy-rules:interaction-factor} of the TOPT
rules in Appendix~\ref{sec:TOPT}, the denominator of each time
interval takes a different form depending upon whether all, some, or
none of the particles are external (e.g. initial or final) particles.
We may factor the fully connected Green's function into parts based
upon the forms of these denominators and their dependence upon the
initial and final relative energies.  We use the superscript
$(4\text{D})$ to denote the part which depends upon both the initial
and the final relative energies, $(4\text{D}\rightarrow3\text{D})$ and
$(3\text{D}\rightarrow4\text{D})$ to denote transitional parts
depending only upon the final or initial relative energies,
respectively, and $(3\text{D})$ to denote the part which is
independent of both the initial and final relative energies.  These
forms are four-dimensional, transitional, and three-dimensional,
respectively.  For the case of $G_c^{(2)}$, these parts are simply
different forms of the two-body interaction.  (The parts are more
complicated for the three body case, as we see later in this Appendix
and in Sec.~\ref{sec:3body-iterative-greens-function}.)  We may see
the relevant differences in these forms of the two-body interaction
most easily by briefly restricting ourselves to the
one-meson-in-flight approximation.  Extension to the complete
interaction is straightforward.

The four-dimensional form of this interaction follows from both
particles being external particles.  Denoting the two-body
center-of-mass energy as $P^0_{12} =  p^0_1 +  p^0_2$, 
\begin{eqnarray}
    V_3^{(4\text{D})} & = &
    \frac{\Gamma_1}
      {\sqrt{2 \omega( \left|{\bf p}_1 - {\bf p}'_1 \right| ) } }
    \frac{1}
      { P^0_{12} - p^0_1 - p'^0_2 - \omega + i\eta}
    \frac{\Gamma_2}
      {\sqrt{2 \omega( \left|{\bf p}_2 - {\bf p}'_2 \right| ) } }
  \nonumber \\* && +
    \frac{\Gamma_2}
      {\sqrt{2 \omega( \left|{\bf p}_2 - {\bf p}'_2 \right| ) } }
    \frac{1}
      { P^0_{12} - p'^0_1 - p^0_2 - \omega + i\eta}
    \frac{\Gamma_1}
      {\sqrt{2 \omega( \left|{\bf p}_1 - {\bf p}'_1 \right| ) } }
    \; . 
  \label{eq:v3-4D}
\end{eqnarray}
Note that the denominator depends upon either the initial or final
energies of the particles $p^0$, rather than the ``on-shell'' energy
$\epsilon$.  This is due to the fact that they are always in either
initial or final states, and allows for the full four-dimensional
nature of ${\cal G}$.  This only appears in the Born term, in which
the two particles interact only once.

The transitional form of the potential has two variants: one initial
particle with one internal particle, and one final particle with one
internal particle.  Here we show the second variant explicitly
\begin{eqnarray}
    V_3^{(4\text{D}\rightarrow3\text{D})} & = &
    \frac{\Gamma_1}
      {\sqrt{2 \omega( \left|{\bf q} - {\bf p}'_1 \right| ) } }
    \frac{1}
      { P^0_{12} - \epsilon_1 - p'^0_2 - \omega + i\eta}
    \frac{\Gamma_2}
      {\sqrt{2 \omega( \left|{\bf P}-{\bf q} - {\bf p}'_2 \right| ) } }
  \nonumber \\* && +
    \frac{\Gamma_2}
      {\sqrt{2 \omega( \left|{\bf P}-{\bf q} - {\bf p}'_2 \right| ) } }
    \frac{1}
      { P^0_{12} - p'^0_1 - \epsilon_2 - \omega + i\eta}
    \frac{\Gamma_1}
      {\sqrt{2 \omega( \left|{\bf q} - {\bf p}'_1 \right| ) } }
    \; . 
  \label{eq:v3-4D-to-3D}
\end{eqnarray}
Note that each denominator depends upon one particle's on-shell energy
$\epsilon$, and the other particle's final energy $p^0$.  This is due
to one particle going into its final state at the beginning of the
interaction, while the other does not do so until the end of the
interaction.  The on-shell (internal) particle reflects our
integration over the internal time variables, while the ``off-shell''
(final) particle reflects the (fully specified) four-dimensional
nature of ${\cal G}$.  The other variant of this form
$V_3^{(3\text{D}\rightarrow4\text{D})}$ is similar, and involves the
initial particle states.

Finally we have the three-dimensional form, where all of the particles
are internal particles, and hence we have integrated out their energy
dependence.
\begin{eqnarray}
    V_3^{(3\text{D})} & = &
    \frac{\Gamma_1}
      {\sqrt{2 \omega( \left|{\bf q} - {\bf q}' \right| ) } }
    \frac{1}
      { P^0_{12} - \epsilon_1 - \epsilon'_2 - \omega + i\eta}
    \frac{\Gamma_2}
      {\sqrt{2 \omega( \left|{\bf q}' - {\bf q} \right| ) } }
  \nonumber \\* && +
    \frac{\Gamma_2}
      {\sqrt{2 \omega( \left|{\bf q}' - {\bf q} \right| ) } }
    \frac{1}
      { P^0_{12} - \epsilon'_1 - \epsilon_2 - \omega + i\eta}
    \frac{\Gamma_1}
      {\sqrt{2 \omega( \left|{\bf q} - {\bf q}' \right| ) } }
    \; . 
  \label{eq:v3-3D}
\end{eqnarray}
Note that all of the particles are on shell, as none of them are
initial or final particles.  This reflects our integration over all
internal time variables.  This is the form which appears in the
two-body bound-state equation, and is independent of the initial and
final relative energies.

We have only been considering the one-meson-in-flight approximation,
but these comments hold for the general case of $V_3$.

We also define the two-body global propagator for cluster $l$ 
\[
  G_0^l \equiv 
  \frac{1}{ P^0_{jk} - \epsilon_j -\epsilon_k +i\eta }
  \; ,
\] 
where $P^0_{jk} = p^0 _{j} + p^0 _k $ is the total energy of
the pair.

It is useful to define the right and left two-body wave operators for
cluster $j$
\begin{mathletters}%
\label{eq:wave-operators}
  \begin{eqnarray}
      \Omega^L_j &\equiv& 
      \sum_{m=0}^\infty 
      \left(
        V_j^{(3\text{D})} N^+_k N^+_l 
        G_0^j
      \right)^m
      \; ,
    \\
      \Omega^R_j &\equiv& 
      \sum_{m=0}^\infty 
      \left(
        G_0^j
        V_j^{(3\text{D})} N^+_k N^+_l 
      \right)^m
      \; .
  \end{eqnarray}
\end{mathletters}%
These wave operators transform the three-dimensional two-body free
propagator $G_0^j$ into the full three-dimensional two-body Green's
function $G_2^j$:
\begin{eqnarray}
  G_2^j&=&G_0^j \Omega^L_j= \Omega^R_j G_0^j
  \nonumber \\ 
  &=& G_0^j  + G_0^j V_j^{(3\text{D})} N^+_k N^+_l G_2^j \; .
\label{eq:app-twobody-G}
\end{eqnarray}

\subsubsection{All three particles interacting}

Finally, we have all three particles interacting.  As in the two-body
case [Eq.~(\ref{eq:G12-omega})], we may separate out the Born terms
from the fully connected Greens function.  In this context Born
terms are defined as those in which all three-body reducible time
intervals contain either an initial or final particle; there is no
factor of $G_0$, the three-dimensional three-body-reducible time
interval.
%
\begin{figure}
  \centerline{ \epsfbox{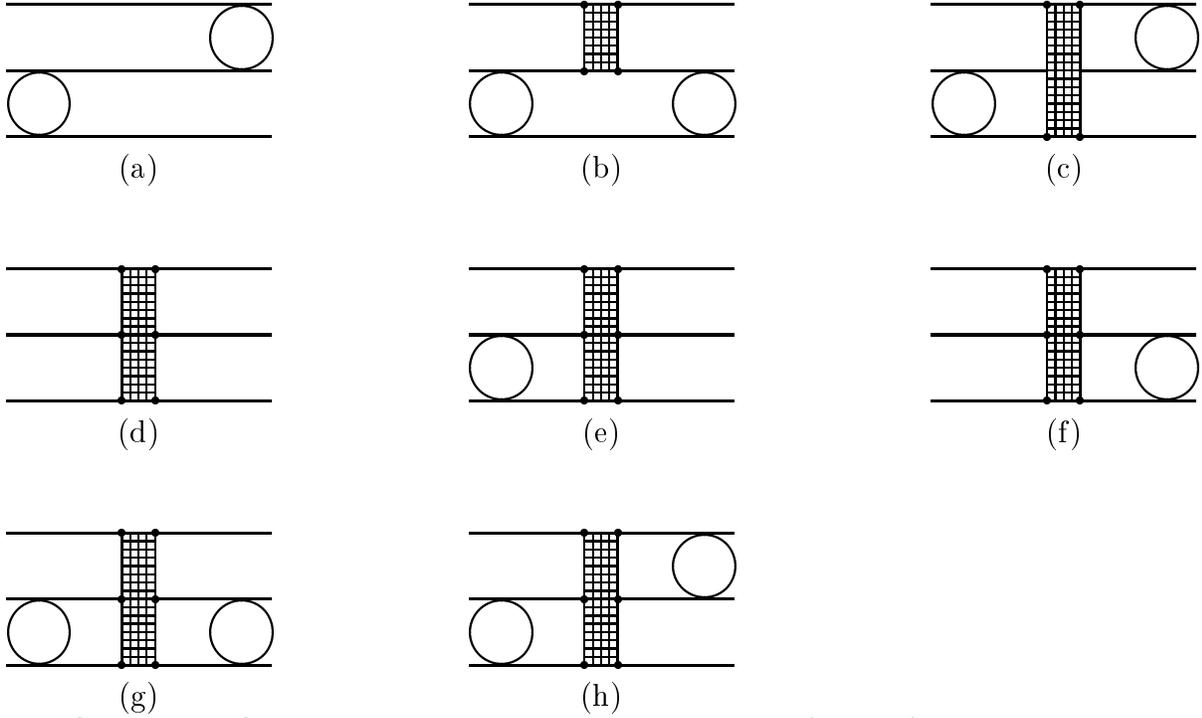} }
  \caption{%
    The TOPT graphs representing the `Born' terms for the
    fully-connected three-body Green's function.  The open circles
    represent one or more two-body forces, the boxes in graphs (b) and
    (c) represent two-body forces, and the boxes in graphs (d) through
    (h) represent three-body forces.  The identifying feature is that
    no graphs contain a three-body reducible time interval that has no
    initial or final particles.
  \label{fig:3Body-Born-terms}
  }
\end{figure}
%
Examples of the three-body Born diagrams are given in
Fig.~\ref{fig:3Body-Born-terms}.  These graphs are the analog of the
two-body $V_3^{(4\text{D})}$.  The simplest example is one of the
terms associated with the graph in Fig.~\ref{fig:3Body-Born-terms}(d);
a single TOPT three-body force.  Choosing the term analogous to that
shown in the second diagram in column (bA) of
Fig.~\ref{fig:reclassify}, but with external legs, we have
\begin{eqnarray}
  \lefteqn{
    V_0^{(4\text{D})} \rightarrow
    i (2\pi)^4 \delta^{(4)}(P - P') \;
    \frac{ N^+_1({\bf p}'_1) }{ p'^0_1 - \epsilon_1({\bf p}'_1) + i\eta} \;
    \frac{ N^+_2({\bf p}'_2) }{ p'^0_2 - \epsilon_2({\bf p}'_2) + i\eta} \;
    \frac{ N^+_3({\bf p}'_3) }{ p'^0_3 - \epsilon_3({\bf p}'_3) + i\eta}
  }
  \nonumber \\* &\qquad& \times
    \frac{ \Gamma_2 N^+_2( {\bf P} - {\bf p}'_1 - {\bf p}_3 ) }
      { \sqrt{2 \omega({\bf p}_3 - {\bf p}'_3)} } \;
    \frac{1}{ P^0 - p'^0_1 - \epsilon_2({\bf P} - {\bf p}'_1 - {\bf p}_3)
      - p'^0_3 - \omega({\bf p}_3 - {\bf p}'_3) + i\eta}
  \nonumber \\* && \phantom{\times} \times
    \frac{ \Gamma_2 N^+_2( {\bf p}_2 ) }
      { \sqrt{2 \omega({\bf p}_1 - {\bf p}'_1)} } \;
    \frac{1}{ P^0 - p'^0_1 - p^0_2 - p'^0_3 
      - \omega({\bf p}_1 - {\bf p}'_1)
      - \omega({\bf p}_3 - {\bf p}'_3) + i\eta} \;
    \frac{ \Gamma_3 N^+_3( {\bf p}_3 ) }
      { \sqrt{2 \omega({\bf p}_3 - {\bf p}'_3)} }
  \nonumber \\* && \phantom{\times} \times
    \frac{1}{ P^0 - p'^0_1 - p^0_2 -  p^0_3
      - \omega({\bf p}_1 - {\bf p}'_1) + i\eta} \;
    \frac{ \Gamma_1 N^+_1( {\bf p}_1 ) }
      { \sqrt{2 \omega({\bf p}_1 - {\bf p}'_1)} }
  \nonumber \\* && \times
    \frac{ 1 }{ p^0_1 - \epsilon_1({\bf p}_1) + i\eta} \;
    \frac{ 1 }{ p^0_2 - \epsilon_2({\bf p}_2) + i\eta} \;
    \frac{ 1 }{ p^0_3 - \epsilon_3({\bf p}_3) + i\eta}
  \; .
\end{eqnarray}
Note that both initial and final particle energies are needed: this is
a fully four-dimensional graph.

In Sec.~\ref{sec:3body-iterative-greens-function} of the main text we
show that when one excludes the three-body Born terms, ${\cal
G}^{(3)}_c$ can be factored in a similar way to
Eq.~(\ref{eq:G12-omega}), as shown in Eq.~(\ref{eq:fG3f}), which we
repeat here for convenience 
\begin{equation}
\eqnum{\ref{eq:fG3f}}
  {\cal G}^{(3)}_c = 
  i  
  (2\pi)^4 \delta^{(4)}(P_f-P_i) 
  f_{\text{post}}^{(4\text{D}\rightarrow3\text{D})}
  G_3^{(3\text{D})}
  f_{\text{pre}}^{(3\text{D}\rightarrow4\text{D})}
  \; .
\end{equation}
Here the pre-{} and post-factors
($f_{\text{pre}}^{(3\text{D}\rightarrow4\text{D})}$ and
$f_{\text{post}}^{(4\text{D}\rightarrow3\text{D})}$) correspond to the
transitional factors containing
$V_3^{(3\text{D}\rightarrow4\text{D})}$ and
$V_3^{(4\text{D}\rightarrow3\text{D})}$ in Eq.~(\ref{eq:G12-omega}).
They connect the three-dimensional $G_3$ to the four dimensional
external world.

  It is interesting to see how ${\cal G}^{(3)}_c$ of
  Eq.~(\ref{eq:fG3f}) vanishes in the limit that one particle does not
  interact (i.e., for sufficiently short-range interactions and when
  there are no zero-energy bound states).  If particle $1$, say, does
  not interact then $V_2$, $V_3$, and $V_0$ vanish: by definition
  these potentials involve particle $1$ interacting at least one time.
  This causes both $f_{\text{pre}}$ and $f_{\text{post}}$
  [Eqs.~(\ref{eq:pre-and-post-factors})] to vanish, as they both
  involve factors of two of the four possible interactions ($V_1$,
  $V_2$, $V_3$, and $V_0$), three of which must now vanish.  
  Note that this does not require $G_3$ of Eq.~(\ref{eq:G3}) to
  vanish, merely the factors which multiply it in the definition of
  ${\cal G}^{(3)}_c$.  In fact, even in the absence of any
  interactions $G_3$ has a nonzero value ($G_0$), however it is not
  physically meaningful; if either $f_{\text{pre}}$ or
  $f_{\text{post}}$ vanishes, then the physically meaningful ${\cal
  G}^{(3)}_c$ vanishes.
  Another way to state this is that ${\cal G}^{(3)}_c$ is fully
  connected, and therefore must vanish in the limit that one particle
  does not interact.  In the factorization of ${\cal G}^{(3)}_c$ we
  have introduced, the pre-{} and post-factors ensure connectedness
  and therefore ensure that it vanishes.

\end{document}